\documentclass{svproc}
\usepackage{graphicx}
\usepackage{float}
\usepackage{amsmath}
\usepackage{amssymb}
\usepackage{caption}
\usepackage[htt]{hyphenat}
\usepackage{subfigure}
\usepackage[toc]{appendix}
\usepackage{url}
\usepackage{amsmath}
\usepackage{adjustbox}
\usepackage{array}
\usepackage{booktabs}

\usepackage{url}

\begin{document}
\mainmatter              
\title{AI Ethics on Blockchain: Topic Analysis on Twitter Data for Blockchain Security\thanks{The corresponding author Luyao Zhang is supported by National Science Foundation China on the project entitled “Trust Mechanism Design on Blockchain: An Interdisciplinary Approach of Game Theory, Reinforcement Learning, and Human-AI Interactions.” (Grant No. 12201266). Yihang Fu is supported by the Summer Research Scholar (SRS) program 2022 under Prof. Luyao Zhang’s project titled “Trust Mechanism Design: Blockchain for Social Good” at Duke Kunshan University. Zesen Zhuang is supported by the Social Science Divisional Chair’s Discretionary Fund for undergraduate research in Prof. Luyao Zhang’s related interdisciplinary research courses as a Teaching and Research Assistant at Duke Kunshan University. Both Zesen Zhuang and Luyao Zhang are also with SciEcon CIC, a not-for-profit organization aiming at cultivating interdisciplinary research of both profound insights and practical impacts in the United Kingdom. We thank the anonymous referees at Computing Conference for their professional and thoughtful comments.}}
\titlerunning{AI Ethics on Blockchain}  

%
\author{Yihang Fu\inst{1}
\and Zesen Zhuang\inst{1} 
\and Luyao Zhang\inst{*}\inst{1}}
\authorrunning{Y. Fu, Z. Zhuang, and L. Zhang} %

\institute{Duke Kunshan University, Suzhou, Jiangsu, 215316, China
\\
\email{*corresponding author:lz183@duke.edu \\
Data Science Research Center \& Social Science Division\\
Duke Kunshan University}\\}

\maketitle              

\begin{abstract}
Blockchain has empowered computer systems to be more secure using a distributed network. However, the current blockchain design suffers from fairness issues in transaction ordering. Miners are able to reorder transactions to generate profits, the so-called miner extractable value (MEV). Existing research recognizes MEV as a severe security issue and proposes potential solutions, including prominent Flashbots. However, previous studies have mostly analyzed blockchain data, which might not capture the impacts of MEV in a much broader AI society. Thus, in this research, we applied natural language processing (NLP) methods to comprehensively analyze topics in tweets on MEV. We collected more than 20000 tweets with \#MEV and \#Flashbots hashtags and analyzed their topics. Our results show that the tweets discussed profound topics of ethical concern, including security, equity, emotional sentiments, and the desire for solutions to MEV. We also identify the co-movements of MEV activities on blockchain and social media platforms. Our study contributes to the literature at the interface of blockchain security, MEV solutions, and AI ethics.

\end{abstract}

\section{Introduction}

Blockchain has enabled computer systems to be more secure using a distributed network.~\cite{zhang2022sok,zhang2022blockchain,zhang2022BNS,ao2022} However, the current blockchain design suffers from fairness issues in transaction ordering.~\cite{kelkar2022order} Miner extractable value (MEV), first coined by Daian et al. in 2020~\cite{daian_flash_2019-1}, refers to the value that miners can extract by reordering the transactions on the blockchain. For example, on the Proof-of-Work (PoW) Ethereum~\cite{ethereum.org}, miners can order, include, and exclude transactions in mem-pool, a pool where transactions are stored or sorted temporarily before adding to the new blocks.  Researchers~\cite{torres2021frontrunner} found that from 2015 to 2020, the 199724 frontrunners had cumulative profits of more than 18.4 billion USD. Since the transition of Ethereum from PoW to Proof-of-Stake (PoS), miners no longer have a role in the blockchain protocol. Instead, validators take charge of validating transactions on the blockchain. However, the method of extracting value by manipulating the transaction order still exists. Therefore, people now use MEV as an abbreviation for the maximum extractable value in PoS Ethereum, the so-called Ethereum 2.0. Existing research recognizes MEV as a severe security issue and proposes potential solutions~\cite{Chainlink,yang_2022_sok} including the prominent Flashbots~\cite {weintraub2022flash}. However, previous studies have mostly analyzed blockchain data, which might not capture the impacts of MEV in a much broader AI society. Therefore, we extend the study of MEV from blockchain data to a broader community on social media platforms. Specifically, our study targets two research questions (RQs):

\begin{enumerate}

\item\textbf{RQ1}: What are the main keywords and topics being
discussed in tweets with \#MEV and \#flashbots hashtags,
and what are the connections between those
keywords?

\item\textbf{RQ2}: What are the connections between the MEV activities on blockchain and discussions on social media platforms? 

\end{enumerate}

In this study, we applied natural language processing (NLP) methods to analyze topics in tweets on MEV comprehensively. We queried more than 20000 tweets with \#MEV and \#Flashbots hashtags from 2019 to October 2022. We also included corresponding Google Trend data in the same period for reference and comparison. To explore the connections between the MEV activities on blockchain and discussions on social media platforms, we collected the gross profit data of MEV from Flashbots. Our results show that the tweets discussed profound topics of ethical concern, including security, equity, emotional sentiments, and the desire for solutions to MEV. According to the keyword statistics, the discussion about MEV is highly concentrated on the Ethereum blockchain. The result also indicates that the MEV problem is one of the most urgent problems on the Ethereum blockchain, and practical solutions are highly demanded. In addition to Flashbots, the topics mention several alternative solutions to MEV, but Flashbots appears to be the most promising one. Some potential nontraditional solutions are also mentioned, such as machine learning. Moreover, other nontechnical keywords indicate that people generally express negative emotions toward MEV, e.g., a feeling of unfairness. We also identify the co-movements of MEV activities on blockchain and social media platforms. Our study contributes to the literature at the interface of blockchain security, MEV solutions, and AI ethics. In Section 2, we discuss the related literature and background. Section 3 introduces the data and methodology.  Section 4 presents the results of the two research questions respectively. Section 5 discusses and concludes. We provide a glossary Table~\ref{tab:Glossary Table} in the Appendix.

\begin{figure}
    \centering
    \includegraphics[width=\linewidth]{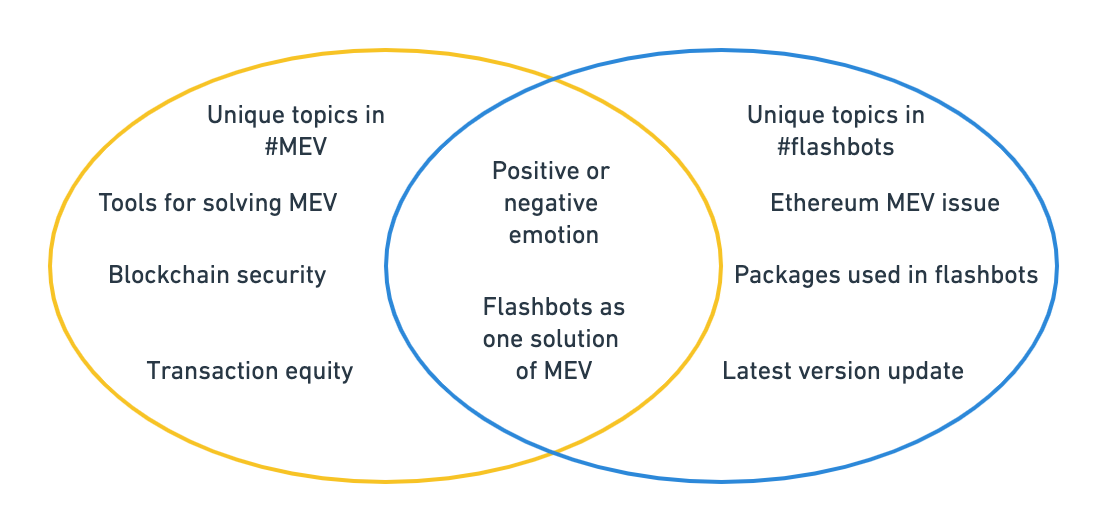}
    \caption{Common topics between MEV and flashbots, and unique
topics in the two hashtags}
    \label{fig:blockchain-security}
\end{figure}

\section{Related Literature and Background}

Our research contributes to three lines of literature: blockchain security, MEV solutions, and AI ethics. 

\subsection{Blockchain Security: MEV issues}

Ethereum blockchain facilitates transactions with the use of smart contracts. In Ethereum, nodes collect transaction information from networks, and miners record the transactions into blocks. Before being added to the blocks, transactions are temporarily stored and sorted in the mem-pool. Miners select transactions in the mem-pool and execute Proof of Work. Whoever (miners) wins the race of PoW can add the block to the network.~\cite{ethereum.org} The order of transactions is predetermined. The execution depends on the initial transaction sets in front of the block or in the same block. However, when Daian et al.~\cite{daian_flash_2019} introduced frontrunning in cryptocurrency decentralized exchange (DEX), miners could change the order of the transactions. In general, MEV is an activity in which attackers (or profit seekers) discover certain instabilities and look for extractable values ~\cite{sandwichattack}. MEV has different strategies to obtain profits. One of the most common strategies is called a sandwich attack. For example, if A is prompting a transaction to purchase a token, the attackers who discovered A’s attempt could buy this token ahead of A at a high price and then sell this token after A. The attackers manage to extract a profit from the series of transactions. Another commonly used strategy is called the arbitrage attack. . In arbitrage, the same good is purchased and sold simultaneously in different markets, and profits are made from the difference in the price of the same good in different markets. In Ethereum, if two or more DEXs offer the same token in different prizes simultaneously, one can buy the cheaper token and sell it at a higher price. Our research contributes to the literature by analyzing concerns about blockchain security discussed on social media platforms. 

\subsection{MEV solutions: Flashbots and alternatives}

Methods to mitigate MEV problems are divided into two main categories: democratization and minimization. MEV minimization sets up roles to make MEV impossible or increases the risk to be larger than the MEV benefits. For example, Ethereum 2.0 upgrades from proof-of-work (PoW) to proof-of-stake (PoS) and introduces slashing ~\cite{piet_extracting_2022} to punish misbehavior regarding MEV. Third-party researchers propose minimization solutions such as fair sequencing services ~\cite{Chainlink} and Conveyor~\cite{network_whats_2021}. Alternatively, MEV democratization tries not to eliminate but to democratize MEV behavior so that everyone has access to information that is available to miners. The most popular approaches, such as Flashbots ~\cite{weintraub2022flash} direct transactions to private relays to reduce the mem-pool bidding war~\cite{noauthor_flashbots_nodate}.\footnote{There are also similar solutions like Eden Network~\cite{piatt_eden_2021} and CoW Protocol~\cite{noauthor_cow_nodate}, etc.} Flashbots aim to mitigate the negative externalities of the current MEV by establishing a fair, transparent, and permissionless ecosystem. Two initiatives mainly support it, \textit{MEV-geth} and \textit{MEV-inspect}~\cite{daian_flash_2019}. Specifically, Flashbots provide a new auction ecosystem with three primary roles: searchers, relays, and miners. Searchers seek MEV opportunities. Once they find potential transactions promoted by peers, they create a bundle that contains a set of transactions. The bundle includes the fee paid to the miners and the searchers themselves. Searchers then send the bundles to the relays instead of the mem pool. Relays collect the bundles and send them to the miners. Since the bundles are sent to the Flashbots, the miners are exclusively Flashbots miners. Miners then collect bundles and select the most profitable ones, and only one transaction can be accounted for in each block. Miners can determine which transaction to mine based on MEV-geth, a forked version of the Go-Ethereum client ~\cite{weintraub2022flash}. Our research contributes to the literature by evaluating MEV solution discussions on social media platforms and their connections to MEV activities on the blockchain. 

\subsection{AI Ethics}

Researchers have expressed concern about the ethical aspects of security issues related to blockchain technologies. Bertino et al. ~\cite{bertino2019data} noted that if data were gathered and used, based on some ethical principles of data transparency, it would provide a novel way for policy-makers to assess the mechanism of blockchain transactions. Another group of researchers proposed a list of ethical concerns about blockchain technology categorized into four areas: technology stack, cryptocurrencies, smart contracts, and decentralization. One of the major concerns in the cryptocurrency area is whether the coin mining mechanism is ethically sustainable and fair ~\cite{tang2019ethics}.
Regarding this question, Ben and his colleagues assessed Flashbots. They argued that in the new auction mechanism of Flashbots (as introduced previously), only some users can receive fair profits, and miners benefit more than searchers~\cite{weintraub2022flash}. However, although the researchers provided comprehensive insights into the ethical discussion, the existing research needs to better evaluate the ethical issue of blockchain based on the real-life reactions of blockchain users. In this article, we measure and evaluate people’s reactions and feedback toward blockchain security issues on social media platforms.

\section{Data and Methodology}

The data and code in this project are open-sourced and can be accessed at \url{https://github.com/SciEcon/blockchain-ethics}
\begin{table}[!htbp]
\centering
\begin{tabular}{|>{\hspace{0pt}}m{0.069\linewidth}|>{\hspace{0pt}}m{0.298\linewidth}|>{\hspace{0pt}}m{0.571\linewidth}|} 
\hline
\textbf{Index} & \textbf{Date} & \textbf{Tweets} \\ 
\hline
0 & 2021-12-31 15:53:38+00:00 & @willwarren No worries @foldfinance solves thi... \\ 
\hline
1 & 2021-12-31 05:56:23+00:00 & Vshadow textbackslash{}n\#Imgoodgirl\textbackslash{}n... \\ 
\hline
2 & 2021-12-31 05:49:28+00:00 & This is what a sandwich attack looks like. The... \\ 
\hline
3 & 2021-12-29 21:11:16+00:00 & \#Memoria...líderes de jxc apoyaron públicament... \\ 
\hline
4 & 2021-12-29 17:34:14+00:00 & even if you don’t have that much money to clai... \\
\hline
\end{tabular}
\caption{Sample Tweets Data}
\label{tab:sample-tweets-data}
\end{table}
\subsection{Data}
We collected three datasets. The first includes tweets and Google Trends data. For Tweets, we used \texttt{snscrape}\footnote{\url{https://github.com/JustAnotherArchivist/snscrape}} to query primary data for our research. \texttt{snscrape} is a Python library to scrape posts on a variety of social networks with specific topics or hashtags. We queried two datasets, one with the hashtag \#mev and the other with the hashtag \#flashbots. We queried Twitter data from 2019-01-01 to 2022-10-01. In total, we found in total 20574 tweets with hashtag \#mev and 852 tweets with hashtag \#flashbots. The queried data includes two columns which are date and content. Table \ref{tab:sample-tweets-data} shows examples of downloaded data. Next, we use Python library \texttt{pytrend}\footnote{\url{https://github.com/GeneralMills/pytrends}} to query Google Trend data for two topics, ``MEV'' and ``flashbots''. pretend provides API to automatically download reports from Google Trend\footnote{\url{https://trends.google.com/trends}}. Then, we merge the Google Trend data with the tweets by date as in Table~\ref{tab:sample-merge-data}. In addition, we also queried Ethereum MEV records from the flashbots' MEV-Explore dashboard\footnote{Flashbots MEV-Explore public dashboard https://explore.flashbots.net/ consists of various historical statistics of MEV activities on the Ethereum blockchain.} to compare on-chain and social media activities.
\begin{table}[!htbp]
\centering
\begin{tabular}{|l|l|l|p{.2\linewidth}|l|} 
\hline
           & \textbf{date} & \textbf{google trend} & \textbf{tweet\_volume} & \textbf{tweet\_len}  \\ 
\hline
\textbf{0} & 2021-04-11    & 23                    & 3                      & 20.666667            \\ 
\hline
\textbf{1} & 2021-04-18    & 23                    & 2                      & 37.500000            \\ 
\hline
\textbf{2} & 2021-04-25    & 0                     & 1                      & 42.000000            \\ 
\hline
\textbf{3} & 2021-05-02    & 0                     & 1                      & 34.000000            \\ 
\hline
\textbf{4} & 2021-05-09    & 24                    & 1                      & 19.000000            \\
\hline
\end{tabular}
\caption{Sample Merged Data: column \textbf{date} shows the date in YYYY-MM-DD format; column \textbf{google trend} shows the Google Trend index; column \textbf{tweet\_volume} is the count of tweets with a specific topic (``MEV'' for example); column \textbf{tweet\_len} is the sum of the length of the tweets in one day.}
\label{tab:sample-merge-data}
\end{table}
\subsection{Methodology}
Our NLP methods include keyword analysis and Latent Dirichlet Allocation (LDA), similar to the quantitative methods in~\cite{tong_what_2022-1}.

\subsubsection{Keywords Analysis Methods}

Analyzing the trend of discussion on a topic on social media and its high relevance is helpful to understand the history and development of the topic and future trends. In this study, we trace and then quantify the activity of the hashtags \#mev and \#flashbots on Twitter and compare them with their Google Trends profiles. We first conduct the Spearman correlation test between tweet volume and Google Trends data and then plot the time series for each hashtag to reveal the correlation between their activity on Twitter and Google. Next, we count and sort the keywords’ appearance (irrelevant words such as emojis and common words are excluded) in tweets and draw a word cloud that shows the most relevant topics discussed on social media to the two hashtags \#mev and \#flashbots. After that, we use Python library \texttt{NetworkX}\footnote{\url{https://networkx.org/}} to draw a network on keywords. The edge in the network indicates that two keywords occur in the same topic, and the thickness of the edge is proportional to the frequency of co-occurrence. 

\subsubsection{Latent Dirichlet Allocation for Topic Analysis}

We utilize Latent Dirichlet Allocation (LDA)~\cite{blei_latent_2003} to reveal the topic tendency of the collected tweets on \#mev and \#flashbots. LDA is a statistical model that groups data and explains why some groups of data are similar. The LDA model is widely used in natural language processing. Our research utilizes an LDA model implemented in the Python library \texttt{gensim}~\cite{rehurek_lrec}. The LDA in \texttt{gensim} implementation has three hyperparameters: (1) integer $K$, the number of topics; (2) rational number $\alpha$ between 0 and 1 controls the per-document topic distribution, a higher $\alpha$ results in more topics in a document; (3) rational number $\beta$ between 0 and 1 controls the per-topic word distribution, a higher $\beta$ results in more keywords in a topic. For results, the LDA model produces the probability of the corpus as shown in equation \eqref{eq:lda}.

\begin{equation}
    \begin{aligned}
    &p(\mathcal{D}|\alpha, \beta)\\
    &=\prod_{d=1}^{K}{
        \int{p(\theta_d|\alpha)}\left(
            \prod_{n=1}^{N_d}{
                \sum_{Z_{d_n}}{
                    p(z_{d_n} | \theta_d)p(w_{d_n}|z_{d_n},\beta)
                }
            }
        \right)d\theta_d
    }
    \label{eq:lda}
    \end{aligned}
\end{equation}

In equation \eqref{eq:lda}, $\theta$ is the joint distribution of a topic mixture, $z$ is the number of topics, and $N$ is the number of words in a set. The model is trained with various $\alpha$, $\beta$, and $K$. We use a coherence score for parameter optimization. A high coherence score in a topic indicates a higher semantic similarity among keywords in which. We manually tried out $K\in\{1,5,10,15,20,25,30\}$ then adopt SA-LDA~\cite{pathik_simulated_2020} algorithm for $\alpha$ and $\beta$ to optimize hyperparameters. Ultimately, we achieved $K=20$, $\alpha=0.31$ and $\beta=0.61$ with a coherence score of 0.4296 for hashtag \#flashbots and $K=5$, $\alpha=0.25$ and $\beta=0.91$ with a coherence score of 0.4687 for hashtag \#mev.

\section{Results}

\subsection{Answers to RQ1}
This section answers the RQ1 based on the four analyses below:
\begin{enumerate}
\item We calculate and ranked the frequent keywords among the tweets under the hashtag of \#MEV and \#Flashbots. 
\item We use the LDA analysis to analyze people's reactions and emotions behind the potential MEV security issue.
\item We establish Network Analysis (NA) to seek the intrinsic relationship between keywords under each Twitter hashtag.
\item We compare the Google Trend data with real-world events.
\end{enumerate}
\subsubsection{Keywords}

\begin{table}[!htbp]
\begin{minipage}[c]{0.5\textwidth}
\begin{tabular}{|l|l|}
\hline
Keywords          & Frequency \\ \hline
\#flashbots       & 513       \\ \hline
\#Flashbots       & 340       \\ \hline
MEV               & 219       \\ \hline
\#MEV             & 152       \\ \hline
ETH               & 85        \\ \hline
mist              & 84        \\ \hline
Flashbots         & 76        \\ \hline
opensea           & 72        \\ \hline
\#Ethereum        & 67        \\ \hline
Support           & 65        \\ \hline
artist            & 65        \\ \hline
grow              & 65        \\ \hline
Shill             & 65        \\ \hline
Shizzlebotz       & 65        \\ \hline
\#nftshill        & 65        \\ \hline
\#PolygonNFT      & 65        \\ \hline
\#openseaNFT      & 65        \\ \hline
\#mev             & 49        \\ \hline
thegostep         & 49        \\ \hline
transactions      & 47        \\ \hline
gas               & 46        \\ \hline
miners            & 41        \\ \hline
MIST              & 41        \\ \hline
\#DeFi            & 40        \\ \hline
Ethereum          & 39        \\ \hline
team              & 39        \\ \hline
bertcmiller       & 31        \\ \hline
transaction       & 31        \\ \hline
\#FlashBots       & 29        \\ \hline
\#mistX           & 28        \\ \hline
NFT               & 26        \\ \hline
front             & 25        \\ \hline
\#riverfrontrocks & 25        \\ \hline
\#ethereum        & 24        \\ \hline
EST               & 24        \\ \hline
\end{tabular}
\caption{Wordcount for \#flashbots}
\label{wordcount_flashbot}
\end{minipage}
\begin{minipage}[c]{0.5\textwidth}
\begin{tabular}{|l|l|}
\hline
Keywords           & Frequency \\ \hline
\#MEV              & 19928     \\ \hline
\#arbitrage        & 7511      \\ \hline
Ecocent            & 5966      \\ \hline
MEV                & 5957      \\ \hline
WETH               & 5339      \\ \hline
video              & 4398      \\ \hline
simply             & 4357      \\ \hline
explained          & 4352      \\ \hline
System             & 4351      \\ \hline
\#HotWater         & 4349      \\ \hline
info               & 3930      \\ \hline
USDT               & 3318      \\ \hline
USDC               & 3290      \\ \hline
ROI                & 2908      \\ \hline
ESP                & 2749      \\ \hline
view               & 2578      \\ \hline
WBNB               & 2382      \\ \hline
triangular         & 2102      \\ \hline
sandwich           & 2078      \\ \hline
spatial            & 1570      \\ \hline
profit             & 1267      \\ \hline
DAI                & 1225      \\ \hline
days               & 1175      \\ \hline
contract           & 1075      \\ \hline
eye                & 1016      \\ \hline
\#SandwichAttacker & 1009      \\ \hline
SAP                & 910       \\ \hline
EigenPhi           & 877       \\ \hline
\#mev              & 816       \\ \hline
pass               & 809       \\ \hline
WBTC               & 768       \\ \hline
BUSD               & 754       \\ \hline
\end{tabular}
\caption{Wordcount for \#MEV}
\label{wordcount_MEV}
\end{minipage}
\end{table}

\begin{figure}[!htbp]
    \centering
    \includegraphics[width=0.8\linewidth]{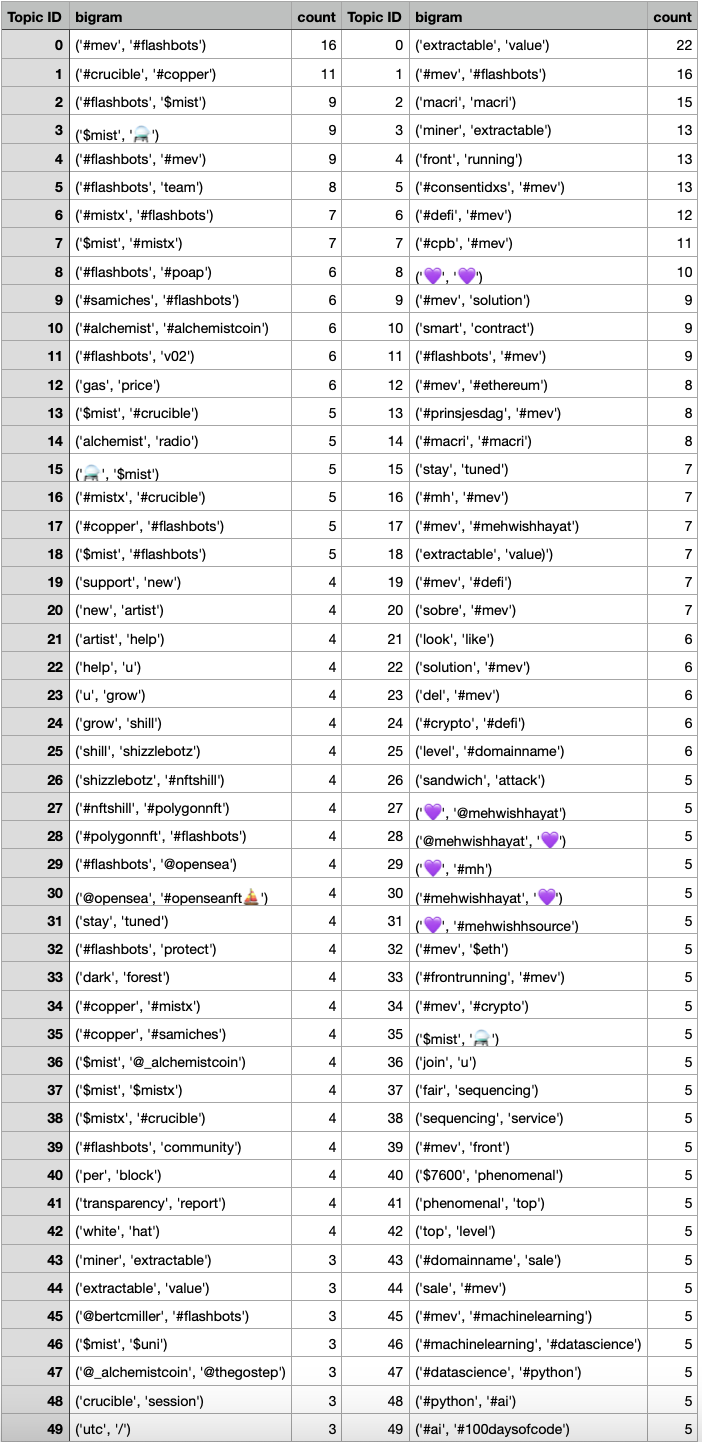}
    \caption{Bigram of keywords: the bigram of keywords, which illustrates the most frequent keyword pairs that were mentioned in tweets under the hashtag of \#MEV and \#Flashbots.}
    \label{Bigram}
\end{figure}

We rank the frequency of keywords for both hashtags as in Table~\ref{wordcount_flashbot} and Table~\ref{wordcount_MEV}. By calculating the word count of the keywords in each topic, we found similarities between these two hashtags. First, both have hashtags with relative semantic meanings. For example, like \textit{Ethereum}, \textit{miner}, and \textit{crypto}, the terminologies in blockchain, are the most salient keywords, appearing more than 300 times, except for \#MEV and \#Flashbots. Also, we observed that \#MEV or \#Flashbots would be mentioned in the other hashtag simultaneously. This reflects the coherent relationship between these two hashtags. As we can see through the table~\ref{wordcount_MEV} and \ref{wordcount_flashbot}, Flashbots was mentioned under the topic of MEV nearly 400 times, and MEV was mentioned nearly 400 times as well under the topic of Flashbots. After investigating the word frequency, we also explored the keyword connections by building up the word bigrams. As we can see in figure \ref{Bigram}, with the selection of the semantic meaning, the top frequent bigrams for the \#flashbots are "mev" \& "flashbots", "crucible" \& "copper", and "flashbots" \& "mist". The top frequent bigrams for the \#MEV are "extractable" \& "value", "mev" \& "flashbots", "macri" \& "macri", "miner" \& "extractable", "front" \& "running", "defi" \& "mev" and "cpb" \& "mev".
\subsubsection{LDA analysis}

\begin{figure}[!htbp]
    \centering
    \includegraphics[width=\linewidth]{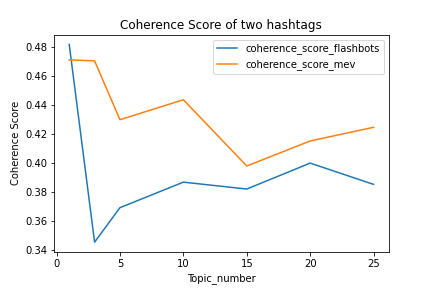}
    \caption{Coherence Score for \#MEV}
    \label{Coherence score}
\end{figure}

We used Genism, an open-source library for unsupervised topic modeling to implement text analysis. We selected 1, 3, 5, 10, 15, 20, 25 as the targeted number of topics to calculate the corresponding coherence score. We found that when k = 1 (the number of topics), the coherence score of \#flashbots (0.482) is the highest. When k = 3, the coherence score of \#MEV, which equals 0.47 is the highest outcome in our model. This indicates that when the number of topics for \#flashbots is 1, and the number of topics for \#MEV is 3, the words in the corpus are relatively more semantically interpretable to humans. Figure \ref{Coherence score} plots the different coherence scores under the different number of topics for two hashtags. Since the coherence score is also attributable to the two
hyperparameters, we adopted Pathik's SA-LDA algorithm to find a pair of approximately suitable $\alpha$ and $\beta$. We found that when $\alpha$ = 0.31, $\beta$ = 0.61, and the number of topics = 20, the coherence value of \#flashbots equals 0.4296, which is the highest among all the outcomes of tested combinations. In the same way, when $\alpha$= 0.25, $\beta$ = 0.91, and the topic = 5, the approximate highest coherence value of \#MEV equals to 0.4687. After excluding some words without semantic meaning, for example, emojis and persons’ names, we identify the sets of words categorized by different topics in Table~\ref{Semantic}. We summarize our main findings in Figure~\ref{fig:blockchain-security}. Our results show that the tweets discussed profound topics of ethical concerns including security, equity, emotional sentiments, and craving for solutions of MEV. Table \ref{Salient terms} illustrates the top 30 salient terms generated by the LDA model with the selected parameters of \#MEV and \#flashbots. Combined with LDA analysis, we find that the top 30 salient terms, including some frequently mentioned keywords such as \texttt{ethereum}, \texttt{smart contract}, and \texttt{solution}, etc., are not categorized by the topics. That is to say, the LDA model did not recognize some of their semantic meanings successfully.

\begin{table}[!htbp]
    \centering
\begin{tabular}{ |p{3cm}||p{3cm}|}
\hline
\multicolumn{2}{|c|}{Top 30 Most Relevant Terms (Overall term frequency)} \\
\hline
\#Flashbots& \#MEV\\
\hline
mist&macri\\
mistx&mev\\
flashbots&extractable\\
crucible&ethereum\\
gt&value\\
gas&see\\
bundles&op\\
copper&chain\\
team&mist\\
poap&inflate\\
alchemist&si\\
eth&aiz\\
samiches&eth\\
mev&flashbots\\
going&video\\
pm&door\\
bertcmiller&mehwishhayat\\
good&chainlink\\
future&energy\\
week&mempool\\
thanks&front\\
see&capital\\
ethereum&love\\
thegostep&miner\\
crypto&us\\
leaksblockchain&worldpastsaday\\
new&look\\
block&people\\
via&makes\\
miners&fair\\
\hline
\end{tabular}
\caption{Top 30 most salient terms of two hashtags}
\label{Salient terms}
\end{table}

\subsubsection{Keywords Network}
\begin{figure}[!htbp]
    \centering
    \begin{subfigure}
      \centering
      \includegraphics[width=0.7\linewidth]{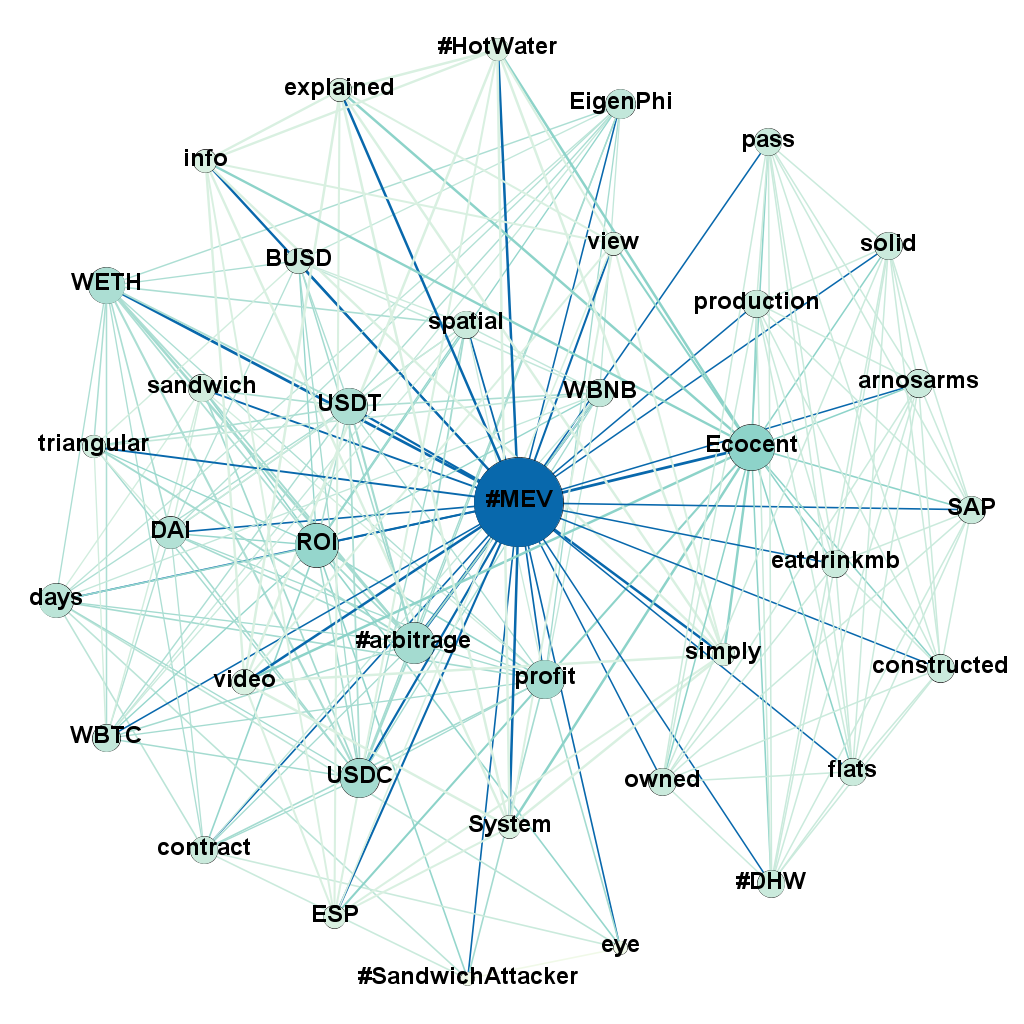}
      \caption{Network of keywords for \#MEV}
      \label{Network1}
    \end{subfigure}
    \par\bigskip 
    \begin{subfigure}
       \centering
       \includegraphics[width=0.7\linewidth]{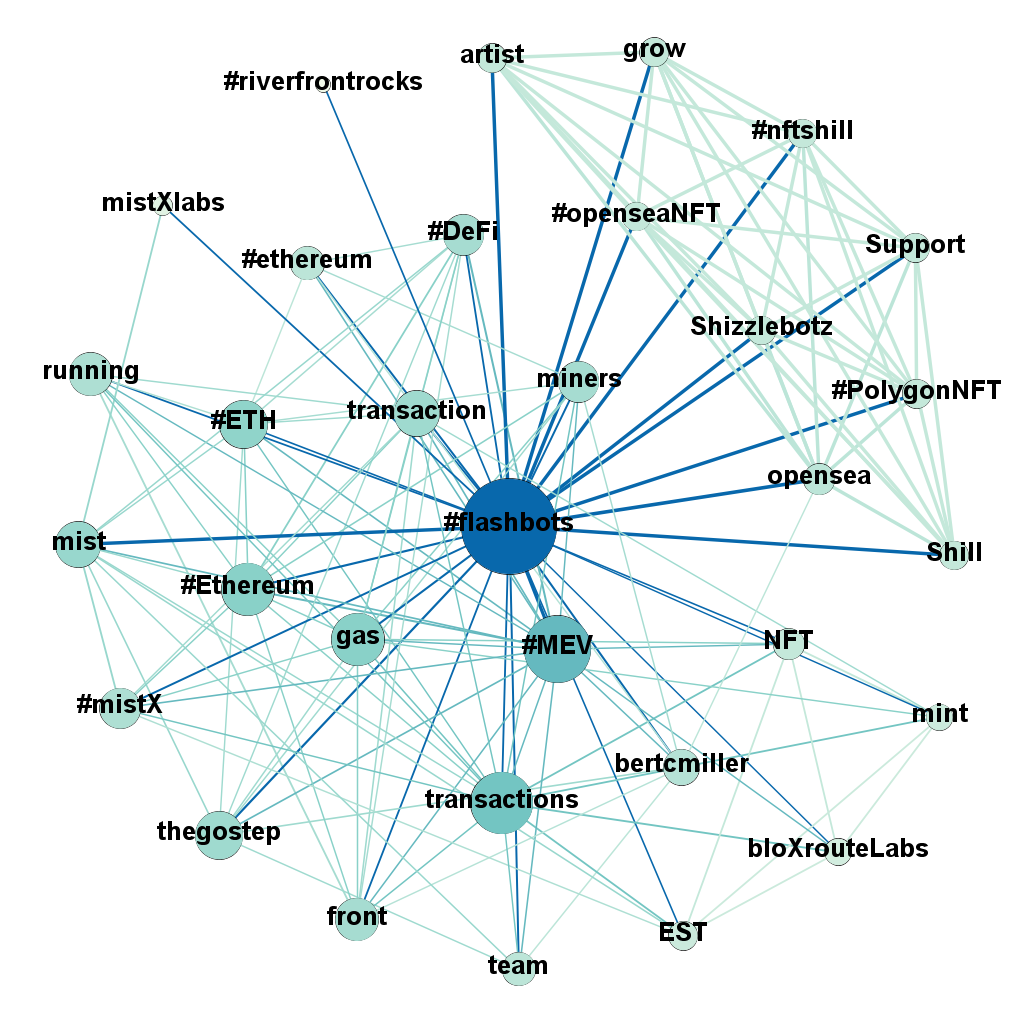}
       \caption{Network of keywords for \#flashbots}
       \label{Network2}
    \end{subfigure}
    
\end{figure}

We used NetworkX on Python to visualize the outcome found in Figure~\ref{Network1} and~\ref{Network2}. Each node represents a commonly used keyword extracted from tweets of \#MEV and \#flashbots, and each edge represents a connection between two keywords. The number of edges of each node represents the number of respective networks with other keywords. We can identify the most commonly mentioned keywords and their relative co-occurring keywords by this analysis. We can see in the solution that the most frequent keywords that appeared together with \#flashbots are \texttt{ethereum}, \texttt{smartcontract}, \texttt{solution}, etc.  The most frequent keywords connected with \#MEV are  \texttt{flashbots}, \texttt{sandwich}, \texttt{miner}, etc.

\subsubsection{Google Trend and Twitter Data}
We also researched Google Trends data of two keywords (namely, \#MEV and \#flashbots) to compare their Twitter volume and offline activities. We used the Google Trends API Pytrend to query the Google Trends data. Google Trends reveals the popularity of top search queries in Google Search across various regions and languages. In general, the Google Trends of \#MEV and \#flashbots show moderate consistency with the respective Twitter volumes. We ran the Spearman correlation test between Twitter volume and Google Trends, and the results showed that \#MEV had more moderate consistency (with coefficient= 0.45) than \#flashbots (with coefficient = 0.202).

\begin{figure}
    \centering
    \includegraphics[width=\linewidth]{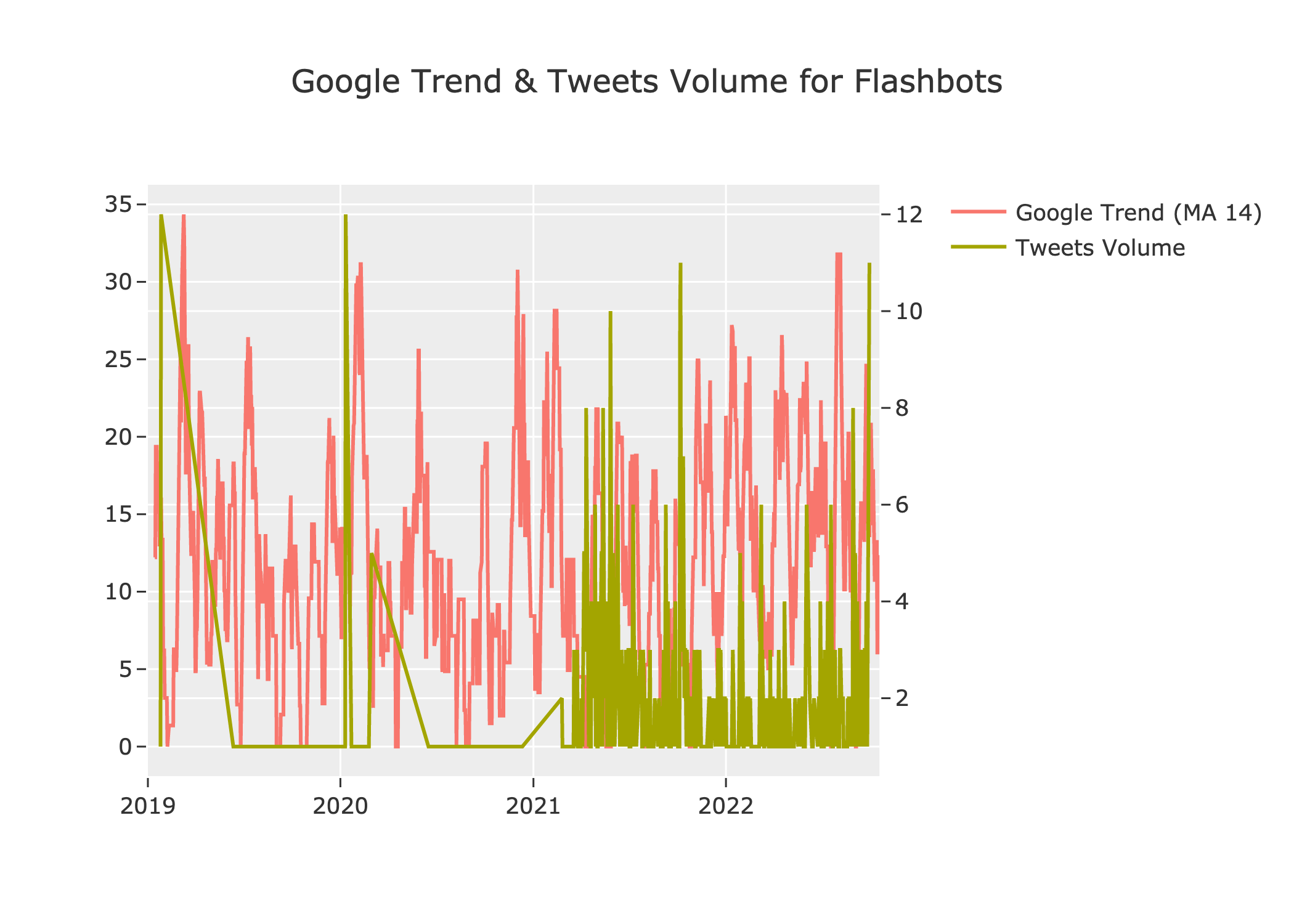}
    \caption{Time series for Google trend and Twitter volume of \#Flashbots}
    \label{Timeseries_flashbots}
\end{figure}

\begin{figure}
    \centering
    \includegraphics[width=\linewidth]{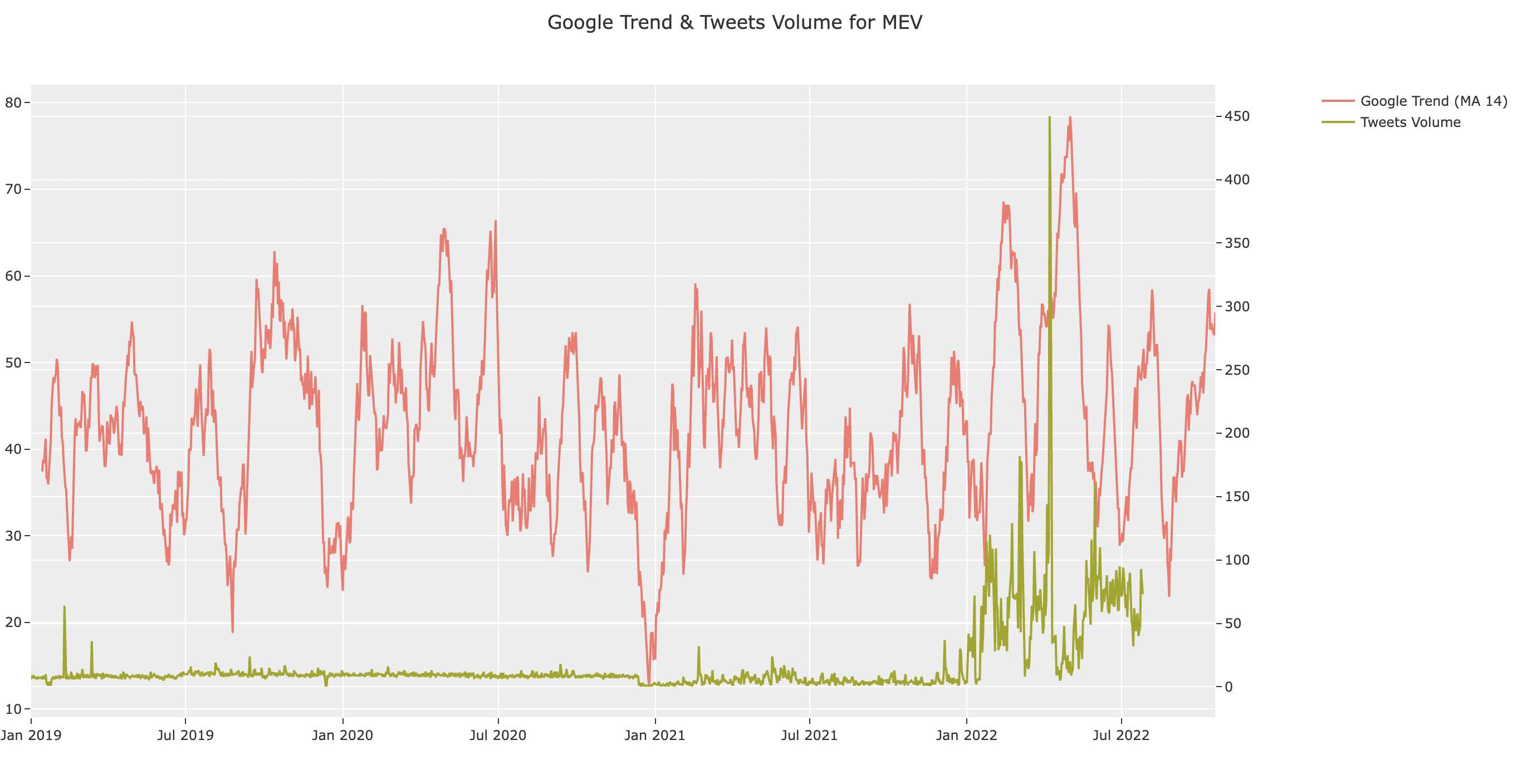}
    \caption{Time series for Google trend and Twitter volume of \#MEV}
    \label{Timeseries_MEV}
\end{figure}

 Furthermore, we observed that each peak of flashbots
google trends have a certain time interval, with the peaks of Flashbots twitter volume. Figure~\ref{Timeseries_flashbots} illustrates the staggering peaks of Google Trend and Twitter volume for \#Flashbots. Also, we found that every peak of Google Trend or Twitter volume of \#Flashbots could match up with a big offline event of Flashbots corporation. For example, on January 2021, Flashbots Auction Alpha (v0.1) was made available for miners and searchers to adopt. The same year, in May, August, and September, Flashbots Auction Alpha updated and published its latest versions corresponding to the peaks shown in Figure~\ref{Timeseries_MEV}. This will be further discussed in the discussion part. However, the situation of \#MEV was slightly different.
Although the Spearman correlation test indicates a higher correlation between Google trends and Twitter volume, those two lines do not show much consistency in peaks. 
\subsection{Answers to RQ2}
\begin{figure}
    \centering
    \includegraphics[width=\linewidth]{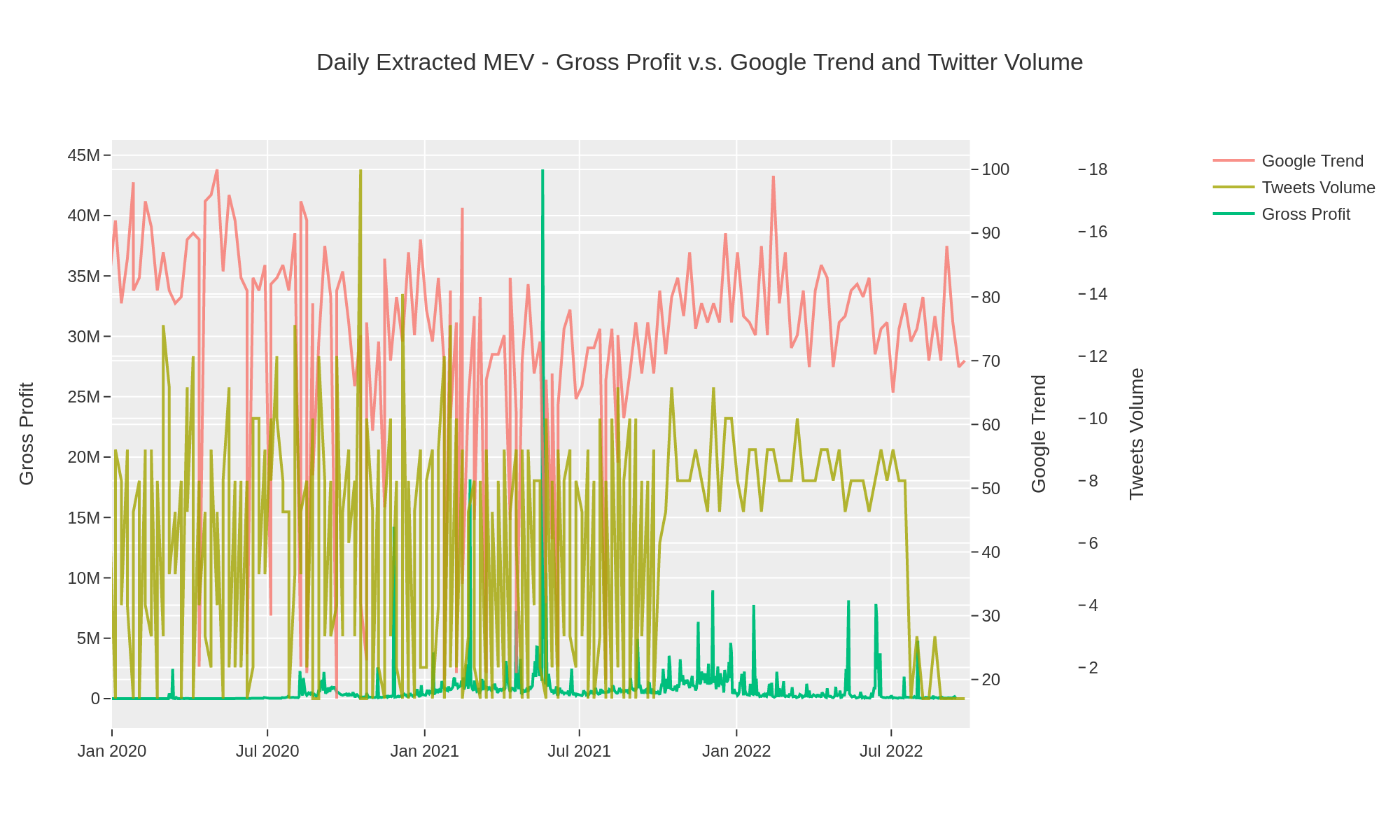}
    \caption{MEV Gross profit, tweet volume, and Google trend }
    \label{Grossprofit}
\end{figure}

To respond to this research question, we query the gross profit data of MEV from 2019 to 2022 using Flashbots API. Gross profit here refers to the amount that attackers acquire from MEV arbitrages. As presented in figure~\ref{Grossprofit}, we observe that there exist three spikes between late 2020 and July 2021. At the same time, we compare the gross profit data with Twitter volume data. Interestingly, we find that each spike in gross profit data corresponds to a spike in Twitter volume
data.  We find that After July 2021, the gross profit of MEV remained at a relatively stable and low value when the Twitter volume was at a high level. Thus the high Twitter volume could not be explained by the gross profit in MEV. Instead, We discovered some potential causes from offline events. In the third quarter of 2021, EIP-1559~\cite{liu2022}, a proposal to reform the Ethereum fee market, and a new version of Zero-Knowledge Rollups, ZK-rollups~\cite{ZK-rollups}, regarded as one of the complete solutions to prevent MEV problems,  were released. We calculated the frequency of the keywords before and after July of 2021 separately, and we found the frequency of keywords "Solution", "Prevent" and "Attack" after 2021.7 was higher than the time before that time. Thus, the release of ZK-rollups and the new Ethereum transaction mechanism might be the driving force for the high social media appearance of MEV topics. 
\section{Conclusion and Discussion}
In conclusion, we apply Natural Language Processing (NLP) methods to comprehensively analyze topics in tweets of MEV.  Our results show that the tweets discussed profound topics of ethical concerns including security, equity, emotional sentiments, and the craving for solutions of MEV. We also identify the co-movements of MEV activities on blockchain and on social media platforms. Our study contributes to the literature at the interface of blockchain security, MEV solutions, and AI ethics. 

\begin{figure}
    \centering
    \includegraphics[width=\linewidth]{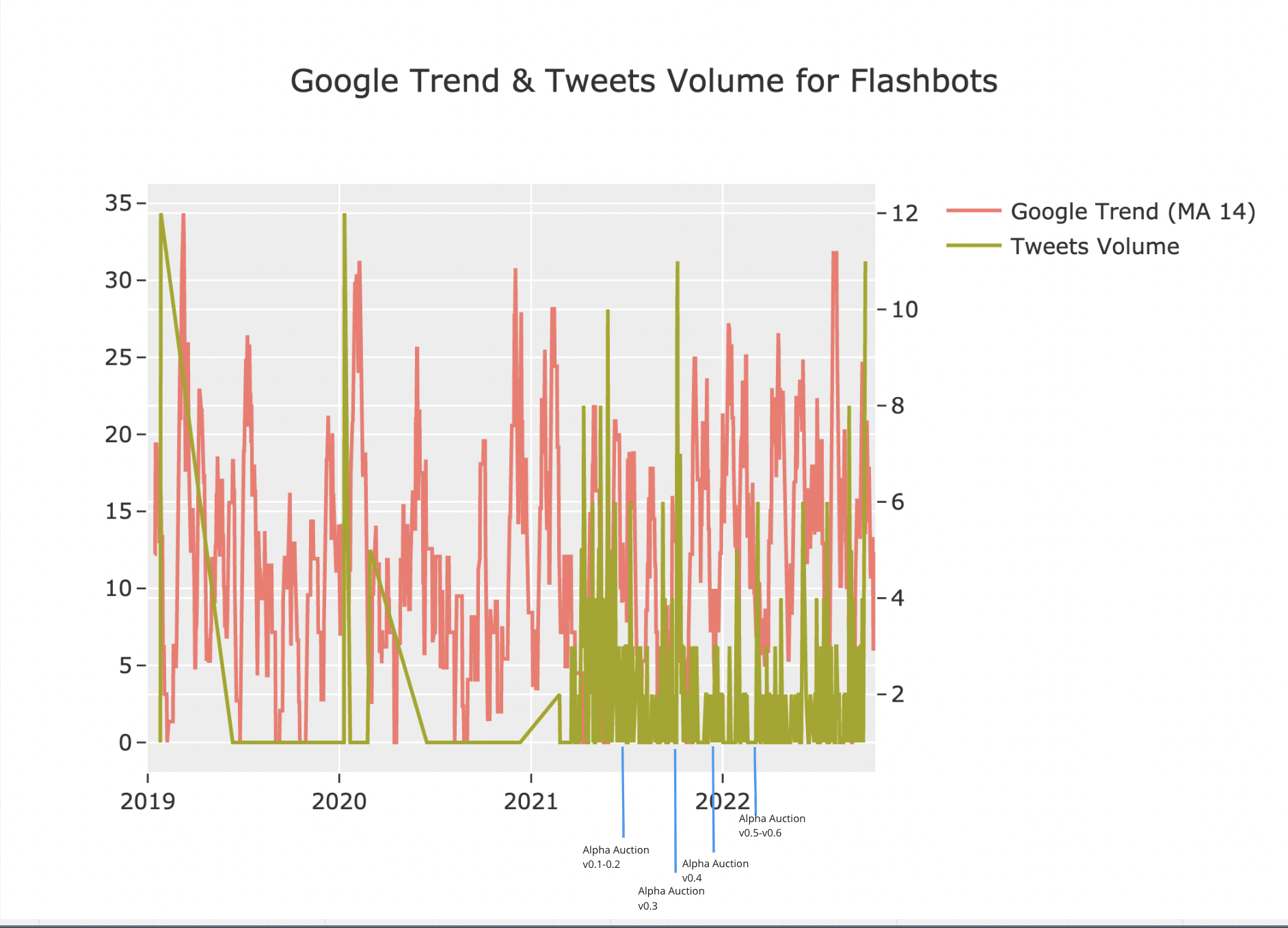}
    \caption{\#Flashbots time-series analysis with landmark events}
    \label{timeseries_withlabels}
\end{figure}

Regarding the different peak times of Google trend and Twitter volume of \#Flashbots, we identify connections between the peaks and the new version release in figure~\ref{timeseries_withlabels}. We observed that the announcements of every latest version of Flashbots will be released through Twitter. Therefore, the peaks of the Google trend obviously tend to appear right after the peaks of Twitter volume. The increases in Google and Twitter trends around the new Flashbots version releases are likely to be from the core teams of Flashbots. Future research could further explore how the discussions of \#mev and \#flashbots differ between users of different backgrounds and changes upon blockchain mechanism upgrades~\cite{zhang2023understand}. For example, how are the topics differ between the core developer team and a broader community of the general public? 
In the bigrams and the networks in Figure~\ref{Bigram}, for \#MEV, the keywords that appear simultaneously with MEV (or mev), such as \texttt{ethereum}, \texttt{machinelearning}, and \texttt{solution}, indicate that MEV happened more often in Ethereum Blockchain; for \#Flashbots, frequent keywords such as frontrunning and sandwich explain the core problems solved by Flashbots, namely, the sandwich attacks. Further research could study how the topics differ in alternative MEV solutions other than Flashbots. 

Table~\ref{Semantic} shows the successfully categorized topics by LDA. In \#flashbots, the most frequently discussed topics are about some blockchain terminologies and DEXs platforms. Another salient topic under this hashtag expresses people's emotional and ethical sentiments, such as fairness, trust, gratefulness, expectation, etc. In contrast, the topics under hashtag \#MEV show people's negative concerns about blockchain security issues such as inflation, unfairness, etc. However, existing research points out that the LDA model has a problem with processing sentiment analysis in short text~\cite{wu2021sentiment}. Therefore, we will consider ameliorating the model in future studies.
\begin{table}[!htbp]
\centering
\begin{tabular}{@{}lll@{}}
\toprule
  & \#MEV                & \#Flashbots               \\ \midrule
1 & Terminologies        & Terminologies             \\
2 & Platforms, companies & Platforms, companies      \\
3 & Concerns, Worries    & Mechanisms, auction, etc. \\
4 & Fairness             & Trust, grateful, etc.     \\
5 & Complaints           & Future, wishes            \\ \bottomrule
\end{tabular}
\caption{Semantic category of the most salient keywords}
\label{Semantic}
\end{table}
\bibliographystyle{spmpsci}
\bibliography{Refs}

\begin{thebibliography}{10}
\providecommand{\url}[1]{{#1}}
\providecommand{\urlprefix}{URL }
\expandafter\ifx\csname urlstyle\endcsname\relax
  \providecommand{\doi}[1]{DOI~\discretionary{}{}{}#1}\else
  \providecommand{\doi}{DOI~\discretionary{}{}{}\begingroup
  \urlstyle{rm}\Url}\fi

\bibitem{sandwichattack}
\emph{Sandwich attack} (2021).
\newblock \url{https://www.mev.wiki/attack-examples/sandwich-attack} [Accessed:
  Whenever]

\bibitem{ZK-rollups}
\emph{MEV-resistant ZK-Rollups with Practical VDE (PVDE)} (2022).
\newblock
  \url{https://ethresear.ch/t/mev-resistant-zk-rollups-with-practical-vde-pvde/12677}
  [Accessed: Whenever]

\bibitem{ao2022}
Ao, Z., Horvath, G., Zhang, L.: Are decentralized finance really decentralized?
  {A} social network analysis of the {Aave} protocol on the {Ethereum}
  blockchain.
\newblock arXiv preprint arXiv:2206.08401  (2022).
\newblock \doi{10.48550/arXiv.2206.08401}.
\newblock \urlprefix\url{https://arxiv.org/abs/2206.08401}

\bibitem{bertino2019data}
Bertino, E., Kundu, A., Sura, Z.: Data transparency with blockchain and ai
  ethics.
\newblock Journal of Data and Information Quality (JDIQ) \textbf{11}(4), 1--8
  (2019)

\bibitem{blei_latent_2003}
Blei, D.M., Ng, A.Y., Jordan, M.I.: Latent dirichlet allocation.
\newblock Journal of machine Learning research \textbf{3}(Jan), 993--1022
  (2003)

\bibitem{daian_flash_2019-1}
Daian, P., Goldfeder, S., Kell, T., Li, Y., Zhao, X., Bentov, I., Breidenbach,
  L., Juels, A.: Flash {Boys} 2.0: {Frontrunning}, {Transaction} {Reordering},
  and {Consensus} {Instability} in {Decentralized} {Exchanges}.
\newblock arXiv:1904.05234 [cs]  (2019).
\newblock \urlprefix\url{https://arxiv.org/abs/1904.05234}

\bibitem{daian_flash_2019}
Daian, P., Goldfeder, S., Kell, T., Li, Y., Zhao, X., Bentov, I., Breidenbach,
  L., Juels, A.: Flash {Boys} 2.0: {Frontrunning}, {Transaction} {Reordering},
  and {Consensus} {Instability} in {Decentralized} {Exchanges}.
\newblock arXiv:1904.05234 [cs]  (2019).
\newblock \urlprefix\url{https://arxiv.org/abs/1904.05234}

\bibitem{Chainlink}
Juels, A.: \emph{Fair Sequencing Services: Enabling a Provably Fair DeFi
  Ecosystem} (2020).
\newblock
  \url{https://blog.chain.link/chainlink-fair-sequencing-services-enabling-a-provably-fair-defi-ecosystem/}
  [Accessed: Whenever]

\bibitem{kelkar2022order}
Kelkar, M., Deb, S., Kannan, S.: Order-fair consensus in the permissionless
  setting.
\newblock In: Proceedings of the 9th ACM on ASIA Public-Key Cryptography
  Workshop, pp. 3--14 (2022)

\bibitem{liu2022}
Liu, Y., Lu, Y., Nayak, K., Zhang, F., Zhang, L., Zhao, Y.: Empirical analysis
  of eip-1559: Transaction fees, waiting times, and consensus security.
\newblock In: Proceedings of the 2022 ACM SIGSAC Conference on Computer and
  Communications Security, CCS '22, p. 2099–2113. Association for Computing
  Machinery, New York, NY, USA (2022).
\newblock \doi{10.1145/3548606.3559341}.
\newblock \urlprefix\url{https://arxiv.org/abs/2305.02552}

\bibitem{nakamoto_bitcoin_2008}
Nakamoto, S.: Bitcoin: {A} peer-to-peer electronic cash system.
\newblock Decentralized Business Review p. 21260 (2008)

\bibitem{network_whats_2021}
Network, A.: What’s {Automata} ({IV}): {Conveyor} (2021).
\newblock
  \urlprefix\url{https://medium.com/atanetwork/whats-automata-iv-conveyor-93c9335e4f43}

\bibitem{pathik_simulated_2020}
Pathik, N., Shukla, P.: Simulated annealing based algorithm for tuning {LDA}
  hyper parameters.
\newblock In: Soft {Computing}: {Theories} and {Applications}, pp. 515--521.
  Springer (2020)

\bibitem{piatt_eden_2021}
Piatt, C., Quesnelle, J., Sheridan, C.: {EDEN} {Network} {Whitepaper} (2021).
\newblock
  \urlprefix\url{https://edennetwork.io/EDEN_Network___Whitepaper___2021_07.pdf}

\bibitem{piet_extracting_2022}
Piet, J., Fairoze, J., Weaver, N.: Extracting {Godl} [sic] from the {Salt}
  {Mines}: {Ethereum} {Miners} {Extracting} {Value} (2022).
\newblock \doi{10.48550/arXiv.2203.15930}.
\newblock \urlprefix\url{http://arxiv.org/abs/2203.15930}.
\newblock ArXiv:2203.15930 [cs]

\bibitem{noauthor_cow_nodate}
Protocol, C.: {CoW} {Protocol} {Overview} (n.d.).
\newblock \urlprefix\url{https://docs.cow.fi/}

\bibitem{rehurek_lrec}
{\v R}eh{\r u}{\v r}ek, R., Sojka, P.: {Software Framework for Topic Modelling
  with Large Corpora}.
\newblock In: {Proceedings of the LREC 2010 Workshop on New Challenges for NLP
  Frameworks}, pp. 45--50. ELRA, Valletta, Malta (2010).
\newblock \url{http://is.muni.cz/publication/884893/en}

\bibitem{ethereum.org}
Smith, C.: \emph{Maximal extractable value (MeV)} (2022).
\newblock \url{https://ethereum.org/en/developers/docs/mev/} [Accessed:
  Whenever]

\bibitem{tang2019ethics}
Tang, Y., Xiong, J., Becerril-Arreola, R., Iyer, L.: Ethics of blockchain: a
  framework of technology, applications, impacts, and research directions.
\newblock Information Technology \& People  (2019)

\bibitem{tong_what_2022-1}
Tong, X., Li, Y., Li, J., Bei, R., Zhang, L.: What are {People} {Talking} about
  in \#{BackLivesMatter} and \#{StopAsianHate}?
\newblock Proceedings of the 2022 AAAI/ACM Conference on AI, Ethics, and
  Society  (2022).
\newblock \doi{10.1145/3514094.3534202}

\bibitem{torres2021frontrunner}
Torres, C.F., Camino, R., et~al.: Frontrunner jones and the raiders of the dark
  forest: An empirical study of frontrunning on the ethereum blockchain.
\newblock In: 30th USENIX Security Symposium (USENIX Security 21), pp.
  1343--1359 (2021)

\bibitem{weintraub2022flash}
Weintraub, B., Torres, C.F., Nita-Rotaru, C., State, R.: A flash (bot) in the
  pan: Measuring maximal extractable value in private pools.
\newblock arXiv preprint arXiv:2206.04185  (2022)

\bibitem{noauthor_flashbots_nodate}
Wiki, M.: Flashbots (n.d.).
\newblock \urlprefix\url{https://www.mev.wiki/solutions/faas-or-meva/flashbots}

\bibitem{wu2021sentiment}
Wu, D., Yang, R., Shen, C.: Sentiment word co-occurrence and knowledge pair
  feature extraction based lda short text clustering algorithm.
\newblock Journal of Intelligent Information Systems \textbf{56}(1), 1--23
  (2021)

\bibitem{yang_2022_sok}
Yang, S., Zhang, F., Huang, K., Chen, X., Yang, Y., Zhu, F.: Sok: Mev
  countermeasures: Theory and practice.
\newblock arXiv:2212.05111 [cs]  (2022).
\newblock \urlprefix\url{https://arxiv.org/abs/2212.05111}

\bibitem{zhang2022sok}
Zhang, L., Ma, X., Liu, Y.: Sok: Blockchain decentralization.
\newblock arXiv preprint arXiv:2205.04256  (2022).
\newblock \doi{10.48550/arXiv.2205.04256}.
\newblock \urlprefix\url{https://arxiv.org/abs/2205.04256}

\bibitem{zhang2022blockchain}
Zhang, L., Tian, X.: On blockchain we cooperate: An evolutionary game
  perspective.
\newblock arXiv preprint arXiv:2212.05357  (2022).
\newblock \doi{10.48550/arXiv.2212.05357}.
\newblock \urlprefix\url{https://arxiv.org/abs/2212.05357}

\bibitem{zhang2023understand}
Zhang, L., Zhang, F.: Understand waiting time in transaction fee mechanism: An
  interdisciplinary perspective.
\newblock arXiv preprint arXiv:2305.02552  (2023).
\newblock \doi{10.48550/arXiv.2305.02552}.
\newblock \urlprefix\url{https://arxiv.org/abs/2305.02552}

\bibitem{zhang2022BNS}
Zhang, Y., Chen, Z., Sun, Y., Liu, Y., Zhang, L.: Blockchain network analysis:
  A comparative study of decentralized banks.
\newblock arXiv preprint arXiv:2212.05632  (2022).
\newblock \doi{10.48550/arXiv.2212.05632}.
\newblock \urlprefix\url{https://arxiv.org/abs/2212.05632}

\end{thebibliography}

\newpage
\appendix
\section{Appendix}
\label{appendix}
\begin{table}[!htbp]
    \centering
    \begin{tabular}{|p{.28\linewidth}|p{.38\linewidth}|p{.28\linewidth}|}
    \hline
         Keywords& Definition & Citation\\ 
         \hline
         Blockchain & Blockchain is a form of DLT (Distributed Ledge Technology) that stores data in a chain of blocks. Each block needs to be verified, validated, and then chained to another block &
         ~\cite{nakamoto_bitcoin_2008}\\ 
         \hline
         Ethereum & A blockchain platform & \url{https://ethereum.org/en/developers/docs/intro-to-ethereum/} \\\hline
         DEXs & “Decentralized Exchanges where a smart contract or other forms of peer-to-peer network executes exchange functionality.” & ~\cite{daian_flash_2019} \\ 
         \hline
         Consensus Mechanism & “The entire stack of protocols, incentives, and ideas that allow a network of nodes to agree on the state of a blockchain.” & \url{https://ethereum.org/en/developers/docs/consensus-mechanisms/} \\ 
         \hline
         PoW & Proof of Work is the mechanism that once allowed the decentralized Ethereum network to come to a consensus (i.e. all nodes agree) on things like account balances and the order of transactions. & ~\cite{nakamoto_bitcoin_2008} \\ 
         \hline
         PoW and Mining & Proof of Stake is the current consensus mechanism of Ethereum Blockchain. Ethereum uses proof-of-stake, where validators explicitly stake capital in the form of ETH into a smart contract on Ethereum. & \url{https://ethereum.org/en/developers/docs/consensus-mechanisms/pos/} \\ \hline
         The Merge & The event that Ethereum decided to change the PoW mechanisms to PoS since the latter is more secure and less resource-intensive & \url{https://ethereum.org/en/upgrades/merge/} \\ \hline    
    \end{tabular}
    \caption{The glossary table}
    \label{tab:Glossary Table}
\end{table}
\end{document}